\documentstyle[12pt]{article}

\catcode`\@=11
\long\def\@makefntext#1{
\protect\noindent \hbox to 3.2pt {\hskip-.9pt
$^{{\ninerm\@thefnmark}}$\hfil}#1\hfill}                

 \def\@makefnmark{\hbox to 0pt{$^{\@thefnmark}$\hss}}  

\def\ps@myheadings{\let\@mkboth\@gobbletwo
\def\@oddhead{\hbox{}
\rightmark\hfil\ninerm\thepage}
\def\@oddfoot{}\def\@evenhead{\ninerm\thepage\hfil
\leftmark\hbox{}}\def\@evenfoot{}
\def\sectionmark##1{}\def\subsectionmark##1{}}

\newcounter{sectionc}\newcounter{subsectionc}\newcounter{subsubsectionc}
\renewcommand{\section}[1] {\vspace{0.6cm}\addtocounter{sectionc}{1}
\setcounter{subsectionc}{0}\setcounter{subsubsectionc}{0}\noindent
	{\bf\thesectionc. #1}\par\vspace{0.4cm}}
\renewcommand{\subsection}[1] {\vspace{0.6cm}\addtocounter{subsectionc}{1}
	\setcounter{subsubsectionc}{0}\noindent
	{\it\thesectionc.\thesubsectionc. #1}\par\vspace{0.4cm}}
\renewcommand{\subsubsection}[1]
{\vspace{0.6cm}\addtocounter{subsubsectionc}{1}
	\noindent {\rm\thesectionc.\thesubsectionc.\thesubsubsectionc.
	#1}\par\vspace{0.4cm}}

\newcounter{appendixc}
\newcounter{subappendixc}[appendixc]
\newcounter{subsubappendixc}[subappendixc]

\renewcommand{\appendix}[1] {\vspace{0.6cm}
	\refstepcounter{appendixc}
	\setcounter{figure}{0}
	\setcounter{table}{0}
	\setcounter{equation}{0}
	\renewcommand{\thefigure}{\Alph{appendixc}.\arabic{figure}}
	\renewcommand{\thetable}{\Alph{appendixc}.\arabic{table}}
	\renewcommand{\theappendixc}{\Alph{appendixc}}
	\renewcommand{\theequation}{\Alph{appendixc}.\arabic{equation}}
	\noindent{\bf Appendix \theappendixc #1}\par\vspace{0.4cm}}

\def\abstracts#1{{
	\centering{\begin{minipage}{30pc}\tenrm\baselineskip=12pt\noindent
	\centerline{\tenrm ABSTRACT}\vspace{0.3cm}
	\parindent=0pt #1
	\end{minipage}}\par}}

\renewenvironment{thebibliography}[1]
	{\begin{list}{\arabic{enumi}.}
	{\usecounter{enumi}\setlength{\parsep}{0pt}
\setlength{\leftmargin 1.25cm}{\rightmargin 0pt}
	 \setlength{\itemsep}{0pt} \settowidth
	{\labelwidth}{#1.}\sloppy}}{\end{list}}

\newcounter{itemlistc}
\newcounter{romanlistc}
\newcounter{alphlistc}
\newcounter{arabiclistc}

\newcommand{\fcaption}[1]{
	\refstepcounter{figure}
	\setbox\@tempboxa = \hbox{\tenrm Fig.~\thefigure. #1}
	\ifdim \wd\@tempboxa > 6in
	   {\begin{center}
	\parbox{6in}{\tenrm\baselineskip=12pt Fig.~\thefigure. #1}
	    \end{center}}
	\else
	     {\begin{center}
	     {\tenrm Fig.~\thefigure. #1}
	      \end{center}}
	\fi}

\newcommand{\tcaption}[1]{
	\refstepcounter{table}
	\setbox\@tempboxa = \hbox{\tenrm Table~\thetable. #1}
	\ifdim \wd\@tempboxa > 6in
	   {\begin{center}
	\parbox{6in}{\tenrm\baselineskip=12pt Table~\thetable. #1}
	    \end{center}}
	\else
	     {\begin{center}
	     {\tenrm Table~\thetable. #1}
	      \end{center}}
	\fi}

\def\@citex[#1]#2{\if@filesw\immediate\write\@auxout
	{\string\citation{#2}}\fi
\def\@citea{}\@cite{\@for\@citeb:=#2\do
	{\@citea\def\@citea{,}\@ifundefined
	{b@\@citeb}{{\bf ?}\@warning
	{Citation `\@citeb' on page \thepage \space undefined}}
	{\csname b@\@citeb\endcsname}}}{#1}}

\newif\if@cghi
\def\cite{\@cghitrue\@ifnextchar [{\@tempswatrue
	\@citex}{\@tempswafalse\@citex[]}}
\def\citelow{\@cghifalse\@ifnextchar [{\@tempswatrue
	\@citex}{\@tempswafalse\@citex[]}}
\def\@cite#1#2{{$\null^{#1}$\if@tempswa\typeout
	{IJCGA warning: optional citation argument
	ignored: `#2'} \fi}}

\def\fnt#1#2{\footnotetext{\kern-.3em
	{$^{\mbox{\sevenrm #1}}$}{#2}}}

 1
 1
 1

\font\tenbf=cmbx10
\font\tenrm=cmr10
\font\tenit=cmti10

\font\ninerm=cmr9

\newcommand{\beq}{\begin{equation}}
\newcommand{\eeq}{\end{equation}}
\newcommand{\AL}%
{\mathrel{\raise.5ex\hbox{$<$}\kern-.75em\raise-.5ex\hbox{$\sim$}}}
\newcommand{\AG}%
{\mathrel{\raise.5ex\hbox{$>$}\kern-.75em\raise-.5ex\hbox{$\sim$}}}

\begin{document}
\centerline{\tenbf QUARK OFF-SHELL EFFECTS IN}
\baselineskip=22pt
\centerline{\tenbf FLAVOUR-CHANGING
DECAYS\footnote{Presented by Jan O. Eeg. Dedicated to the
memory of our colleague Roger Decker.}}

\vspace{0.8cm}
\centerline{\tenrm J. O. EEG}
\baselineskip=13pt
\centerline{\tenit Department of Physics,
University of Oslo}
\baselineskip=12pt
\centerline{\tenit N-0316 Oslo, Norway}
\vspace{0.3cm}
\centerline{\tenrm and}
\vspace{0.3cm}
\centerline{\tenrm I. PICEK}
\baselineskip=13pt
\centerline{\tenit   Department of Physics, Faculty of Science, University of
Zagreb}
\baselineskip=12pt
\centerline{\tenit  HR-41000 Zagreb, Croatia}
\vspace{0.9cm}
\abstracts{We discuss some flavour-changing
 effective Lagrangians, obtained from scales above 1 GeV,
 which vanish on the quark mass shell.
Although the effects of such effective Lagrangians are zero
in the limit of vanishing bound-state interactions, we have shown
that they have a  significant  impact on the processes
 $K\rightarrow\gamma\gamma$
and $B_s \rightarrow\gamma\gamma$.}
\vfil
\vspace{0.8cm}
\rm\baselineskip=14pt
\section{Introduction}
Dealing with  weak hadronic decays, one faces the problem of overbridging
the quark world  and the meson world where
a physical process occurs. The analyses starting from
the high-energy side evolved from the
traditional Feynman diagram technique to an
implementation of the operator-product
expansion \cite{glam}.
When studying non-leptonic decays of order $\sim G_F$, and
weak radiative  decays of order $\sim e G_F$ or  $\sim e^2 G_F$,
one normally writes down an effective Lagrangian including
operators that contribute to a given process, and operators
that mix with these under QCD renormalization. Within this
standard procedure, one usually omits operators  containing
 $(i \gamma \cdot D~-~m_q)$,  by appealing to the  equations of
motion (EOM) for quark fields \cite{Coll,OP,PolSi}:
\begin{equation}
 (i \gamma \cdot D - m_q) \rightarrow 0 \; ,
\label{eq:onshell}
\end{equation}
where $D_{\mu}$ is the covariant derivative containing the gluon
and the photon fields.
This procedure corresponds to going on-shell with external
quarks in quark operators. Certainly, quarks are not exactly
on-shell in hadrons, especially not in the octet (would-be
Goldstone) mesons $\pi, K, \eta$.
We shall see that the naive use of (\ref{eq:onshell}) is not correct in
general.
In fact, the bound-state interactions within mesons might be understood as
a change of the (perturbative) equations of motion.
One expects on general grounds that $(i \gamma \cdot D - m_q)$,
instead of having zero effect, is proportional to some binding energy
 or some non-perturbative parameter
characterizing the problem.

 We consider two cases where
an effective Lagrangian containing the factor
$(i \gamma \cdot D - m_q)$
has been  studied \cite{ep93}.
First, we consider  the circumstances under which the renormalized
$s \rightarrow d$ self-energy transition
becomes potentially relevant to
 $K \rightarrow 2\pi$ decays. It was previously  shown
 \cite{Don} that non-zero effects of $s \rightarrow d$ transitions
 might persist.
Then we consider Lagrangians obtained from quark
diagrams for $s \rightarrow d \gamma$ and $s \rightarrow d
\gamma \gamma$
relevant to $K  \rightarrow \gamma \gamma$,
and similar Lagrangians, obtained from quark diagrams for
$b\rightarrow s\gamma$ and $b\rightarrow s\gamma\gamma$, relevant
to the $B_{s}\rightarrow\gamma\gamma$ decay.

 A proved example of an off-shell effect is the Lamb shift $\; - \; $
the tiny  difference between the self-energy of a {\em free} electron
 and the self-energy of an
electron  {\em bound} in the H-atom\cite{lamb}.
 One might expect more significant analogous effects for much
 more strongly bound quarks.
 Still, since one can
hardly speak of the referent free-quark self-energy, one might expect
that there is a better chance of
finding an observable effect in  the flavour-changing,  non-diagonal
$s\rightarrow d$ self-energy transition.

\section{ Effective Lagrangians at quark level}

The total effective strangeness-changing Lagrangian may be written in the form
\begin{equation}
{\cal{L}}(\Delta S =1)  = \sum_{i} C_i Q_i  \; \;  ,
\label{eq:efflag}
\end{equation}
where the $C_i$'s are coefficients containing the effects of
short distances through electroweak loop diagrams, dressed with hard
gluons. The $Q_i$'s are
operators containing light-quark fields ($q = u,d,s$).
We consider two specific pieces of ${\cal{L}}(\Delta S =1)$,
namely ${\cal{L}}_{ds}$ due to the non-diagonal $s \rightarrow d$
self-energy transition and
${\cal{L}}(s \rightarrow d)_{\gamma}$
due to  $s \rightarrow d \gamma$ and $s \rightarrow d \gamma \gamma$
 transitions.

\begin{figure}
\vspace{3cm}
\caption{Self-energy transition  diagrams corresponding to  ${\cal{L}}_{ds}$.}
\end{figure}

The bare unrenormalized self-energy transition (see Fig. 1) is divergent and
non-vanishing on the mass shell.
Since there are no direct $s \rightarrow d$
transitions in the original Lagrangian, the renormalization is carried out so
that $s \rightarrow d$ transitions are absent for
on-shell $s$- and $d$-quarks.
This requirement defines the physical $s,d$- fields
in the presense of weak interactions,
and the effective Lagrangian corresponding to the
 renormalized self-energy transition
takes the form
\begin{equation}
{\cal{L}}^R_{ds} = - A \, \bar{d} (i \gamma \cdot D - m_d)
  (i \gamma \cdot D R + M_R R + M_L L)
(i\gamma \cdot D - m_s) s  \; + h.c.  \; \; ,
\label{eq:rense}
\end{equation}
where $M_L, M_R$ are constants depending on quark masses, and
$A$ contains the result of the loop integration.

In the  pure electroweak case, the CP-conserving part of $A$ is
\cite{Chia} of order $G_{F}m^{2}_{c}/M^{2}_{W}$. However,
the CP-violating part of $A$ has no such
suppression for a $t$-quark with a mass of the same order as
 the $W$-boson or heavier.
 Moreover, adding perturbative QCD to lowest order, in any case one obtains
an unsuppressed contribution $\sim G_{F}\alpha_{s} (\log m^2)^2$, where
$m=m_{c}$ and $m \sim M_{W}$ in the CP-conserving and CP-violating
cases, respectively \cite{shab,GPP,JOE87} .

 If one applies the  EOM  as in (\ref{eq:onshell}), then
${\cal{L}}^R_{ds} \rightarrow 0$.
According to the standard procedure,  one would discard
 contributions  from ${\cal{L}}_{ds}^R$ to physical amplitudes,
such as $K \rightarrow 2\pi$.
However, if (\ref{eq:onshell}) is violated for off-shell bound quarks in
$\pi$ and $K$, physical effects could be obtained, and  one
should explore possible consequences for the $\Delta I = 1/2$ rule
and for $\epsilon'/\epsilon$ in $K \rightarrow 2\pi$ decays.

\begin{figure}
\vspace{7cm}
\caption{Quark diagrams for
 $s \rightarrow d \gamma$ and $s \rightarrow d \gamma \gamma $
corresponding to ${\cal{L}}(s \rightarrow d)_{\gamma}$,
${\cal{L}}_F$ and ${\cal{L}}_{\sigma}$ in eqs. (4) - (7).}
\end{figure}

The evaluation of the loop diagrams for
 $s \rightarrow d \gamma$ and $s \rightarrow d \gamma \gamma$
 transitions \cite{sdgg,YW,HK} (for real photons; see Fig. 2)
without going to the mass shell,
 results in an effective Lagrangian \cite{ep93}
\beq
{\cal{L}}(s\rightarrow d)_{\gamma} \, = \,  B \,
\epsilon^{\mu \nu \lambda \rho}
F_{\mu \nu} \, ( \bar{d}_{L} \; i \stackrel{\leftrightarrow}{D_{\lambda}}
 \gamma_{\rho} s_L )  \; + h.c. \; \; ,
\label{eq:lagg}
\eeq
where $F$ is the electromagnetic field tensor, and
$B \sim e G_F \lambda_{KM}$  depends on the loop integration
($\lambda_{KM}$ is the relevant KM parameter).
 In order to follow the
fate of the off-shell contribution, it is convenient to rewrite
(\ref{eq:lagg}) in the form
\beq
{\cal{L}}(s\rightarrow d)_{\gamma} = {\cal{L}}_F +
{\cal{L}}_{\sigma}  \;  ,
\label{eq:lagfs}
\eeq
where
\beq
{\cal{L}}_F = B_F \, \bar{d} [(i\gamma \cdot D - m_d) \,
 \sigma_{\mu \nu} F^{\mu \nu} L + \sigma_{\mu \nu} F^{\mu \nu} R (i\gamma
\cdot D -m_s)] s  \; + h.c. \; \; ,
\label{eq:lagf}
\eeq
and  ${\cal{L}}_{\sigma}$ is the well-known magnetic-moment term,
\beq
{\cal{L}}_{\sigma} = B_{\sigma}
 \, \bar{d} \, (m_s \sigma_{\mu \nu} F^{\mu \nu} R +
 m_d  \sigma_{\mu \nu} F^{\mu \nu} L) \, s  \;  + h.c. \; \; .
\label{eq:lags}
\eeq
Here we  have anticipated that the coefficients $B_{F}$ and $B_{\sigma}$,
being equal at the $W$-scale, evolve differently down to
the scale of $\sim$ 1 GeV. The anomalous dimension of ${\cal{L}}_F$
is zero \cite{cella}, while  ${\cal{L}}_{\sigma}$ is known to
have a nonzero anomalous dimension \cite{shiea}.

It has been shown that ${\cal{L}}_F$ does  {\em not} contribute to
 $s \rightarrow d \gamma \gamma$  when the external quarks are on-shell:
 The irreducible $s \rightarrow d \gamma \gamma$ part, with $i D_{\mu}
\rightarrow e_{s(d)} A_{\mu}$, is exactly cancelled by reducible
diagrams \cite{YW,HK},
i.e. with one photon on an external line of  the $s \rightarrow d \gamma$
vertex, with $D_{\mu} \rightarrow \partial_{\mu}$.
Thus, for on-shell quarks, the remaining contribution from
${\cal{L}}(s\rightarrow d)_{\gamma}$
to $s \rightarrow d \gamma \gamma$ is due to the reducible diagrams,
where the effective flavour-changing vertex
corresponds to ${\cal{L}}_{\sigma}$ alone. Moreover, this remaining
contribution vanishes in the chiral limit $m_{s,d} \rightarrow 0$,
as seen from (\ref{eq:lags}).
In the pure electroweak  case, the CP-conserving part
of the quantity $B$ is very small,
$\sim e G_F m_c^2/M_W^2$, owing
to an effective GIM cancellation between $u$- and $c$- quarks, whereas
the CP-violating
part is substantial ($\sim e G_F$) owing to
the heavy $t$-quark \cite{sdgg}.
In the CP-conserving case, a significant
amplitude $\sim e G_F \alpha_s \log(m_c^2)$ is induced by perturbative
QCD \cite{shiea}.

\section{Bosonization and the chiral quark model}

One possibility of including  non-perturbative {\em confining and
 chiral-symmetry} aspects of QCD is to use
some version of the chiral quark
model, an effective low-energy QCD model advocated by many
authors \cite{Weinb/Manohar,Biea,PdeR}.
To quote Weinberg \cite{Weinb/Manohar}, such a framework will introduce
``fictitious elementary
particles into the theory, in rough correspondence with the bound
 states" -- pseudoscalar Goldstone
mesons among the degrees of freedom of the constituent quark model.
The chiral quark model includes the ordinary QCD Lagrangian and
adds a term ${\cal{L}}_{\chi}$ that takes care of
chiral-symmetry breaking,
\begin{equation}
  {\cal{L}}_{\chi} = - M (\bar{q}_R \; U q_L +
\bar{q}_L \; U^\dagger q_R) \; ,
\label{eq:defchi}
\end{equation}
where $\bar{q} = (\bar{u},\bar{d},\bar{s})$ and the $3 \times 3$ matrix
$U \equiv \exp\biggl({2i \Pi/ f}\biggr)$
contains the pseudoscalar octet mesons
 $\Pi = \sum_a \pi^a\lambda^a/2 \,$
$(a=1,..,8)$, and
$f$ can be identified with the pion decay constant,
$f = f_\pi = (92.4 \pm 0.2)$ MeV ($=f_K,$ in the chiral limit).
This term,  proportional to the
constituent quark mass $M \sim$ 300 MeV,
 includes the Goldstone meson octet in  a
chiral-invariant way, and provides a  meson-quark coupling
that makes it possible to calculate matrix elements of quark
operators as loop diagrams. In this effective field theory it is
of course no problem to handle  off-shell quarks.
It should be noted that a calculation
based on (\ref{eq:defchi}) reproduces the amplitude for
 $\pi^0 \rightarrow 2 \gamma$, governed by the triangle anomaly.

 The term ${\cal{L}}_{\chi}$ in (\ref{eq:defchi})
 can be transformed into a pure
mass term $- M \bar{{\cal{Q}}} {\cal{Q}}$
for rotated "constituent quark" fields ${\cal{Q}}_{L,R}$:
\begin{equation}
q_L \rightarrow  {\cal{Q}}_L =  \xi  q_L \; ; \;  \; \;
q_R \rightarrow  {\cal{Q}}_R =  \xi^\dagger q_R \;  ;
\; \; \xi \, \cdot \, \xi  = U \; .
\label{eq:qrot}
\end{equation}
Then the meson-quark couplings in this ``rotated" $(R)$ picture  are
transformed into the kinetic
(Dirac) part of the "constituent quark" Lagrangian.
These interactions can be described in terms of vector and axial vector
fields coupled to constituent quark fields
${\cal{Q}}$.
The rotated picture is of course equivalent to the unrotated picture
 defined by (\ref{eq:defchi}). However,
an explicit diagrammatic evaluation gives
 zero result for $\pi^0 \rightarrow 2 \gamma$ in the rotated
picture.  The explanation is that the anomaly
 term \cite{WZ} is contained in the Jacobian of the quark
field rotation in eq. (\ref{eq:qrot}).

 Although the $\pi^{0}$ axial anomaly is not
conventionally termed the off-shell effect, it might be
viewed in this way, because the divergence of the axial current
cannot be reproduced by the classical equations of motion.

\section{ Application to $K \rightarrow \pi\pi$ and
 $K \rightarrow \gamma\gamma$}

Using the meson-quark couplings $\sim M \gamma_5 / f_{\pi}$
obtained from (\ref{eq:defchi}), and effective electroweak transition
 vertices obtained from ${\cal{L}}^R_{ds}$,  ${\cal{L}}_F$ and
 ${\cal{L}}_{\sigma}$, the calculation of weak transions for mesons
 can be performed in terms of quark loops.
An explicit calculation of the contribution to $K\rightarrow  2\pi$
from ${\cal{L}}^R_{ds}$ gives a non-zero but negligible contribution in
the CP-conserving case \cite{ep93,JOE87}. For the CP-violating
case, one might obtain a significant contribution\cite{BEPE}.
Anyway, it should be noted that the $K_L \rightarrow \pi\pi$ amplitude
obtained from ${\cal{L}}^R_{ds}$ is  suppressed by
 $M^2/\Lambda_{\chi}^2 \, $, where
 $\Lambda_{\chi} = 2 \pi f_{\pi} \sqrt{6/N_c}$ is the chiral
 symmetry-breaking scale. Thus, this contribution  disappears in the limit
$M/f_{\pi} \rightarrow 0$, when the meson-quark interactions from
(\ref{eq:defchi}) are switched off.

In our previous work \cite{sdgg} we found a substantial  CP-violating
$K_S \rightarrow \gamma\gamma$ amplitude  from {\em irreducible}
diagrams for  $s \rightarrow d \gamma \gamma$. Owing to Ward
 identities between
$s \rightarrow d \gamma \gamma$ and $s \rightarrow d \gamma$ transitions,
there is a cancellation  between 1PI diagrams for
 $s \rightarrow d \gamma \gamma$
 and reducible diagrams  for  the two-photon emission
(where the 1PI transition
$s \rightarrow d \gamma$ is a building block). However, one
cannot expect this free-quark cancellation to persist in the real world:
the hadronic matrix elements of the reducible graphs
 (see Fig. 3, middle) are of highly non-local
character, whereas the matrix elements of the irreducible
graphs (see Fig. 3, left) are
proportional to a quark current, having a well-known matrix
 element \cite{sdgg}.

\begin{figure}
\vspace{3cm}
\caption{The total $s \rightarrow d \gamma \gamma$
amplitude due to ${\cal{L}}_F$ (represented by a cross within a circle):
 The irreducible contribution (left) is cancelled by the
reducible contribution (middle) for on-shell quarks. The last diagram (right)
originates from  ${\cal{L}}^R_{ds}$.}
\end{figure}

There have been  considerable efforts \cite{ep93,ep94,k3p} devoted to the
study of the {\em direct} $K_{L,S} \rightarrow \gamma\gamma$ amplitudes
induced by the operators (\ref{eq:lagg})--(\ref{eq:lags}).
By explicit calculation within the chiral
quark model, we have found a non-zero contribution to
$K \rightarrow \gamma\gamma$ from ${\cal{L}}_F$. The
result can be written as an effective interaction
\begin{equation}
{\cal L}(K \rightarrow 2\gamma) \, = \,
 G_{K 2 \gamma} \, F \cdot \tilde{F} \, \Phi_K \;  \: \; ;
\;  G_{K 2 \gamma}  \; \sim  \; \;
e \, B_F\, f_{\pi} \, \frac{M^2}{\Lambda_{\chi}^2} \, ,
\label{eq:lagp}
\end{equation}
where $\Phi_K$  is the K-meson field. The result of the
irreducible part of ${\cal{L}}_F$ (Fig. 3, left), being
proportional to $f_K (= f_{\pi}$ in the chiral limit) is cancelled
by the leading part of the reducible graphs (Fig. 3, middle), the
net result being the formally non-leading part of the reducible graphs.
Strictly speaking, (\ref{eq:lagp}) is the result in the chiral limit,
i.e. the contribution from ${\cal L}_F(m_{s,d} \rightarrow 0)
 \, = \, {\cal L}_{\gamma}(s \rightarrow d)$.
It should also be noted that $K \rightarrow \gamma\gamma$ receives a
contribution from ${\cal{L}}_{\sigma}$ (Fig. 3, middle) and
${\cal{L}}^R_{ds}$ (Fig. 3, all diagrams). However, these
contributions are less important numerically.

Although  $G_{K 2 \gamma}$ in (\ref{eq:lagp}) is formally suppressed
 by $M^2/\Lambda_{\chi}^2 \, ,$
 its coefficient is sizeable, yielding
a significant amplitude, both in the
CP-conserving $K_L \rightarrow  \gamma \gamma$ case \cite{ep94}, and in
the CP-violating $K_S \rightarrow  \gamma \gamma$ case \cite{ep93,sdgg}.
Thus we disagree with some authors \cite{HK,Wolf} who claim that this
 effect is unimportant.

An important property of the
$K \rightarrow  \gamma \gamma$ amplitude obtained from
${\cal{L}}_F$, is that it is zero
when diagramatically calculated within
the rotated basis \cite{ep94}.
Thus, this amplitude has a similar anomalous nature as
$\pi^0 \rightarrow  \gamma \gamma$ !
One should, however, note that in contrast to
$\pi^0 \rightarrow  \gamma \gamma$, the contribution
to $K \rightarrow  \gamma \gamma$ from
${\cal{L}}_F$ (and similarly from ${\cal{L}}^R_{ds}$) is
mass-dependent through the  factor $M^2$.

The non-zero contribution from ${\cal{L}}_F$ to
$K \rightarrow  \gamma \gamma$ was recently confirmed in a bound-state
calculation \cite{k3p}.  To evaluate the hadronic matrix elements
in the bound-state approach,
the variant of an effective meson bilocal theory  was used.
 The bound-state  calculation
in essence confirms the   previous chiral-quark results:
our off-shell contribution is, within the language of chiral
perturbation theory, an entirely new  ${\cal{O}}(p^{4})$
{\em direct-decay} piece \cite{ep94} not contained in previous
analysis \cite{eck}, whereas
the {\em reducible} pole contributions \cite{pole} are numerically
uncertain, and the non-diagonal magnetic-moment term belongs
to the  ${\cal{O}}(p^{6})$ terms.

\section{The quark-loop $B_s\rightarrow\gamma\gamma$   amplitude}

As in the $\Delta S = 1$ case \cite{ep93,ep94}, the flavour-changing radiative
vertices have to  be
supplemented by the quark-meson vertex in order to perform the full quark-loop
evaluation. In contradistinction to the analogous
$K \rightarrow \gamma \gamma$ decay \cite{ep93,ep94},
 the heavy $B$ meson cannot be treated as a Goldstone boson of the chiral
quark model adopted earlier. However, the
pseudoscalar character of $B$ mesons allows us to
write down  a simple $b \bar{s} \, B_s$ interaction,
replacing the $\cal{L}_{\chi}$ term in (\ref{eq:defchi}) by
\beq
i G_{B}\bar{s}\gamma_{5} b B_s \; \;  .
\label{eq:qqB}
\eeq

This interaction may in general be  non-local,  i.e.
 $G_B$ might be momentum-dependent \cite{Hold}.
Thereby, as usually done \cite{ep93}, we  trade the meson-quark coupling
$G_{B}$ in
favour of the meson-decay constant $f_B$.
In calculating the contributions from
${\cal{L}}_{F}$ and ${\cal{L}}_{\sigma}$ in (\ref{eq:lagf})
and (\ref{eq:lags}), respectively
(with the obvious replacements $s \rightarrow b$ and $d \rightarrow s$),
 we obtained \cite{ep94B} an amplitude  of the following form
\beq
M(B_{s}\rightarrow \gamma\gamma)  \; = \; e_D \, f_{B}
[ A_{(+)} \, F_{\mu\nu} F^{\mu\nu} +
 i  A_{(-)} \, F_{\mu\nu}\tilde {F}^{\mu\nu}]  \;  ,
\label{eq:mgen}
\eeq
\beq
A_{(\pm)} \; = \;
 \tau_{F}^{(\pm)} B_F + B_{\sigma} \tau_{\sigma}^{(\pm)} \,   \;  ,
\label{eq:mgenp}
\eeq
where the quantities $\tau_{F,\sigma}^{(\pm)}$ are dimensionless and
depend on the bound-state dynamics. Numerically, they turn
 out to be of order one.
 The coefficients
  $B_F$ of ${\cal{L}}_{F}$ and $B_{\sigma}$
of  ${\cal{L}}_{\sigma}$
now contain the KM factors relevant to the $b \rightarrow s$ transition
and are renormalized at the scale $\mu = m_b$.

In the formal limit where the current quark masses $m_{b,s} \rightarrow 0$,
the $F_{\mu\nu}\tilde {F}^{\mu\nu}$ term of (\ref{eq:mgen}) should reduce to
the anomalous $K \rightarrow 2 \gamma$ amplitude described in the
preceding section. However, in the real world,
 $M_b \gg M_s \AG M \sim$  300~MeV, and the  result
for Br($B_s \rightarrow 2 \gamma$) will be rather different from
$K \rightarrow 2 \gamma$. In order to estimate the
model-dependent quantities  $\tau_{F,\sigma}^{(\pm)}$,  one might
 consider several  assumptions.
In a previous paper \cite{ep94B} we found that, with reasonable assumptions,
$\tau_{\sigma}^{(\pm)} \sim$ 2 to 3, whereas
$\tau_{F}^{(\pm)}$ was formally suppressed
by $\tilde{\Lambda}/M_b$ with respect to $\tau_{\sigma}^{(\pm)}$.
(The parameter $\tilde{\Lambda}$ is a few hundred MeV).
This is in agreement with  intuitive expectations.
 We also found that the genuine off-shell
 term ${\cal{L}}_{F}$ increased the rate by a factor  of $\sim$
 1.5 to 3.
We  conclude that  the branching ratios of the order
$10^{-8}$ to $10^{-7}$ are realistic.
  This  result is not far from that given in Ref.\cite{Sing}.
Our prediction is still two orders of magnitude above the LD estimates
based on the vector-meson dominance \cite{tramp}.

In a recent paper \cite{Carls}, it was reported  that there
are off-shell bound state effects in the process
$B \rightarrow K^* \gamma$. However, only the off-shellness of the $b$-quark
was taken into account, whereas we have found that for $B_s \rightarrow 2
\gamma$
 the off-shellness of the $s$-quark
cannot be neglected.
 In our recent study \cite{ep94B} of
$B_s \rightarrow 2\gamma$ we have shown that the contribution
from the two-photon piece of ${\cal{L}}_F$ is exactly cancelled by parts
of its contribution from the one-photon piece. The remaining
contribution from the off-shell operator ${\cal{L}}_F$ corresponds to loop
diagrams containing  effective $B_s \, \bar{b} b \gamma$
and $B_s \, \bar{s} s \gamma$ vertices.
This result is equivalent to that presented in Ref.\cite{ep93}:
${\cal{L}}_F$ may be transformed  into the wave function, but then it
reappears in the bound-state dynamics.

\section{Results and conclusions}

We have demonstrated the quark off-shell effects in flavour-changing
two-photon decays, such as $s\rightarrow d\gamma \gamma$ ($b\rightarrow
 s\gamma\gamma$) and its
hadronic ${\bar K}^{0}\rightarrow\gamma\gamma$ ($B_{s} \rightarrow \gamma
 \gamma$) counterparts. Thus,  the same basic off-shell effect seems to
take place in processes belonging to such different calculating environments
as the chiral perturbation theory and the one accounting for the
heavy-light bound-states.
Thus the naive use of the (perturbative)
EOM (\ref{eq:onshell}) is not applicable in general.
The bound-state dynamics changes  the (perturbative)
equations of motion (\ref{eq:onshell}).

The genuine off-shell effects are formally suppressed
 in a certain limit by $\tilde{\Lambda}/M_b$
 for $B \rightarrow \gamma \gamma$ and by $(M/\Lambda_{\chi})^2$
for $K \rightarrow\gamma\gamma$ (and $K \rightarrow \pi \pi$).
 Numerically, the suppression is not equally
pronounced in these  cases. For $K \rightarrow\gamma\gamma$, the effect
of ${\cal{L}}_{F}$ is bigger than that of  ${\cal{L}}_{\sigma}$. Indeed,
the latter  effect  is  chirally suppressed
and  of order  ${\cal{O}}(p^6)$.

We have assumed that chiral-symmetry breaking only affects
the strong sector, and does not induce any
new terms in the electroweak sector. It might be argued that this
is not obvious\cite{Hans}. It is of course possible to write down
effective Lagrangian terms containing meson fields in the combined
strong and electroweak sector which contribute to our processes.
However, we cannot see how such terms should be generated.
One way of  addressing this issue could be within
 Nambu-type models.
Within such models, one (or more) gluon exchanges generate four
quark operators in the strong sector which are supposed \cite{BRE}
 to be responsible for the
term (\ref{eq:defchi}). Apriori, ${\cal{L}}_F$ could
generate a new relevant operator if another quark line is attached
through gluon exchange. However, the sum of such contributions
vanishes on the mass shell, and  for off-shell quarks, will
correspond to a complicated higher dimensional operator which,
for instance, will not  cancel our
 $K \rightarrow \gamma \gamma$ amplitude.

The quark off-shellness represents a link that brings the electroweak
$K \rightarrow\gamma\gamma$ decays close to the electromagnetic
$\pi^0 \rightarrow \gamma \gamma$ decay \cite{EPFiz}.
 Although the $\pi^{0}$ axial anomaly is not
conventionally termed the off-shell effect, it might be
viewed in this way, because the divergence of the axial current
cannot be reproduced by the classical equations of motion.
It should be emphasized that $K \rightarrow\gamma\gamma$ can be
treated as textbook approaches for $\pi^0 \rightarrow\gamma\gamma$,
 with one of the electromagnetic
vertices replaced by that obtained from ${\cal{L}}_{F}$.
Strictly speaking, all that is assumed through our treatment of
$K \rightarrow 2\gamma$ is the PCAC ! : the K-field is replaced
 by the divergence of the axial quark current (corresponding to
the rotated picture), or by the pseudoscalar quark density
(unrotated picture).

However, the direct amplitude originating in the quark off-shellness
 in the kaon
is only a fraction of the total $K_{L}\rightarrow \gamma \gamma$
amplitude, and it is model-dependent through the constituent
quark mass $M$.
 The various LD aspects, including the reducible pole
contributions, seem to play a dominant role in this case \cite{pole}.
However, the CP-violating $K_{S}\rightarrow \gamma \gamma$
amplitude receives its main contribution from ${\cal{L}}_F$ \cite{sdgg}.

 For $B_{s} \rightarrow \gamma
 \gamma$, we have also found  a non-zero genuine off-shell contribution.
Although the hadronic matrix element is model dependent, there are  substantial
off-shell contributions that increase the rate by a factor  of $\sim$ 1.5 to 3.
It is hoped that some of  the uncertainties in calculating the effects
of ${\cal{L}}_{F}$ could be resolved
within some variant of a QCD sum-rule \cite{Nar} calculation.

\section{Acknowledgments}
I.P. acknowledges the support of the EU contract CI1*--CT91--0893(HSMU).
J.O.E. acknowledges the hospitality of the Dept. of Physics,
University of Zagreb, some parts of the autumn 1994.

\end{document}